# Vortex in liquid films from concentrated surfactant solutions containing micelles and colloidal particles

Elka S. Basheva[a], Peter A. Kralchevsky[a,1], Krassimir D. Danov[a], Rumyana D. Stanimirova[a], Neil Shaw[b], and Jordan T. Petkov[b]

[a] Department of Chemical and Pharmaceutical Engineering, Faculty of Chemistry and Pharmacy, Sofia University, Sofia 1164, Bulgaria

[b] Lonza Arch UK Biocides Ltd., Hexagon Tower, Crumpsall Vale, Blackley, Manchester M9 8GQ, UK

ORCID Identifiers: Peter A. Kralchevsky: 0000-0003-3942-1411 ; Krassimir D. Danov: 0000-0002-9563-0974

**Abstract**

*Hypothesis:* New dynamic phenomena can be observed in evaporating free liquid films from colloidal solutions with bimodal particle size distribution. Such distributions are formed in a natural way in mixed (slightly turbid) solutions of cationic and anionic surfactants, where nanosized micelles coexist with micronsized precipitated particles.

*Experiment*: Without evaporation of water, the films thin down to thickness < 100 nm. Upon water evaporation from the film, one observes spontaneous film thickening (above 300 nm) and appearance of a dynamic vortex with a spot of thinner film in the center. The vortex wall has a stepwise profile with step-height equal to the effective micelle diameter (ca. 8 nm) and up to 20–30 stratification steps.

*Results:* For thicknesses greater than 100 nm, stratification in foam films from micellar solutions has never been observed so far. It evidences for the formation of a thick colloidal crystal of micelles in the evaporating film. The role of the bigger, micronsized particles is to form a filtration cake in the Plateau border, which supports the thick film. The developed quantitative mechanical model shows that the stepwise vortex profile is stabilized by the balance of hydrodynamic and surface tension forces. Vortex is observed not only in films from catanionic surfactant solutions, but also in films from silica and latex particle suspensions, which contain smaller surfactant micelles.

[1] To whom correspondence should be addressed. Email: pk@lcpe.uni-sofia.bg

*Address*: Faculty of Chemistry and Pharmacy, Sofia University,
1, James Bourchier Blvd., Sofia 1164, Bulgaria

# 1. Introduction

If the effective volume fraction of spherical surfactant micelles in an aqueous solution exceeds ca. 15%, one observes stepwise thinning (stratification) of the liquid films formed from this solution, which is due to layer-by-layer thinning of a micellar structure formed in the film [1,2]. This phenomenon has been observed not only with surfactant micelles, but also with various colloidal particles [1,3–7]. It is a manifestation of the oscillatory structural surface force, which is a result of overlap of the structured layers of molecules or colloidal particles in the neighborhood of the film surfaces [8–12]. In the case of electrically charged micelles (particles), the effective volume fraction includes also a contribution from their counterion atmospheres [1,13,14]. In the case of lower particle volume fractions, the oscillatory structural force degenerates to the depletion attraction [9,11,15].

From the height of the stratification steps, one can determine the micelle aggregation number of both nonionic [16] and ionic [13,14] surfactant micelles. In the ideal case of hard spheres, the particle suspension exists in a fluid-like state up to $\phi = 0.494$ particle volume fraction, which corresponds to the Kirkwood–Alder–Wainwright phase transition [17,18]. The range of action of the structural surface force is expected to expand for ordered phases of colloidal particles above the bulk phase transition [19].

Structural forces are observed also in solutions containing elongated (ellipsoidal, rodlike) micelles. If the micelles can freely rotate and the film thickness is sufficiently large, the period of the structural force is determined by the micelle hydrodynamic diameter measured, e.g., by dynamic light scattering [20]. In thinner quasi-equilibrium films and higher micelle concentrations, the steric effect of walls orients the elongated micelles parallel to the film surfaces, so that the oscillatory period is determined by the rod diameter. In draining thinner films, the shear flow creates hydrodynamic torques that tend to rotate the elongated micelles and to turn them aside of parallel orientation, which leads to a dynamic increase of the period of oscillatory force at smaller film thicknesses [20].

The evaporation of solvent (e.g. water) from a liquid film gives rise to a convective flux from the surrounding Plateau border to the film, thus, compensating the evaporated solvent. If the flux carries along particles within the film, they form 2D ordered arrays compacted by the lateral capillary forces [21]. This phenomenon led to the method of convective self-assembly [22], which has found numerous applications for the production of nanostructured surfaces and materials [23–27].

Solvent evaporation is known to give rise also to other interesting dynamic phenomena in liquid films. The "tears of wine" represent a classical example, where the driving forces are surface-tension gradients engendered by the nonhomogeneous evaporation of ethanol dissolved in the aqueous phase [28]. In the case of a pure volatile liquid (no surfactants), the cooling due to evaporation gives rise to thermal gradients of surface tension, which lead to influx of liquid in the film and to its stabilization [29]. In the case of mixtures of liquids with different volatility, the evaporation leads to accumulation of the liquid of lower volatility in the film, which also brings about dynamic film stabilization – the so called solutocapillary effect that has been observed with both foam and wetting films [30,31].

In the present article, for the first time we report the existence of another dynamic phenomenon – vortex in evaporating liquid films formed from solutions, which contain *both* surfactant micelles and bigger colloidal particles. We used mixtures of cationic and anionic surfactants, often called *catanionic* mixtures [32]. They find applications in laundry detergents, softeners and hard surface disinfectants because they combine the cleaning properties of the anionic surfactant with the antimicrobial action of the cationic one [33–35].

Section 2 presents the used materials and methods. In Section 3, the experimental results are reported. In Section 4, the physical origin of the observed phenomena is discussed. In Section 5, a mechanical model is developed, which identifies the forces that support the stepwise profile of the vortex wall. This model allows one to analyze the effects of different factors, such as film thickness, pressure difference, and evaporation rate, on the observed phenomenon. A rich collection of experimental data is presented as Supplementary Information, including three videos.

## 2. Materials and Methods

### 2.1. Materials

In our experiments, the solutions have been prepared by mixing the tri-amine cationic surfactant N-(3-aminopropyl)-N-dodecyl-1,3-propanediamine, known as Lonzabac-12 (LB12), with the anionic surfactant sodium laurylethersulfate with one or two ethylene oxide groups (SLES-1EO or SLES-2EO); see SI Appendix, Fig. S1 (SI = Supplementary Information). LB12 was product of Lonza (Basel). Depending on the pH, LB12 can exist in the form of mono-, di- and tri-valent cation, characterized by $H^+$ association constants [36]: $pK_{a1} = 6.7$; $pK_{a2} = 8.4$, and $pK_{a3} = 10.0$.

SLES-1EO and SLES-2EO, with commercial names STEOL®CS-170 and STEOL®CS-270, were products of Stepan Co., USA. In some of the experiments, we used also the zwitterionic surfactant dodecyl-dimethylamine oxide (DDAO) and the cationic surfactants dodecyl and tetradecyl-ammonium bromide (DTAB and TTAB), all of them products of Sigma Aldrich. Silica particles Excelica UF-305 of mean diameter 2.7 μm produced by Tokuyama Co., Japan, and polystyrene latex particles with sulfate surface groups and mean diameter 1 μm, product of Dow Chem. Co. were also used.

The experimental finding of vortex in foam films from mixed solutions of the tri-amine biocide surfactant LB12 was serendipitous. Next, we established that the phenomenon is more general and can be observed in mixed solutions of common quaternary ammonium surfactants, like DTAB and TTAB, as well as in suspensions containing colloidal particles (silica and latex) plus surfactant micelles; see below.

*2.2. Methods*

The experiments were carried out with individual free foam films formed in Scheludko-Exerowa (SE) cell [37]; see the sketch in Section 4 below. First, the investigated solution is loaded in a cylindrical capillary (of inner radius 1.5 mm) through an orifice in its wall. Thus, a biconcave drop is formed inside the capillary. Next, liquid is sucked through the orifice and the two menisci approach each other until a liquid film is formed in the central part of the cell. By injecting or sucking liquid through the orifice, one can vary the film radius. If the film thickness is smaller than 100 nm, it can be measured in reflected monochromatic light by means of an interferometric method using a photomultiplier [1, 16, 37]. Film thicknesses greater than 100 nm can be estimated from the interference colors obtained in white (polychromatic) light [38,39]; see below. All micrographs were taken by optical Carl Zeiss microscope Axioplan FL, where the source of white light was a halogen bulb. The reference marker of this microscope represents a built-in caliper (Fig. 1), which shows a reference distance of 50 μm in all micrographs presented in this study. The SE cell was placed in a closed container so that the water vapors can be equilibrated with the solution and evaporation from the film can be prevented. Experiments under these conditions are referred as experiments in *closed cell*. If the glass cover of the container is removed, evaporation of water from the film takes place. This is case in which vortex was observed. Measurements under these conditions are referred as experiments in *open cell*. More details can be found in Refs. [1,16,37]. In all experiments, the working temperature was 25 °C.

*Rheological measurements.* The apparent viscosity of the investigated solutions was measured by a rotational rheometer Bohlin Gemini (Malvern Instruments, UK) using cone-and-plate geometry. The cone angle was 2° and the minimal gap distance was 70 μm.

*Particle size distributions* were obtained by dynamic light scattering using Zetasizer Nano ZS (Malvern Instruments, UK).

## 3. Results

*3.1. Description of the vortex in foam films*

Fig. 1 illustrates the stages of evolution of a free foam film formed from an aqueous solution of 400 mM 3:2 SLES-1EO + LB12 at pH = 6. At this pH, the solution is slightly turbid because of the presence of precipitated particles. Its viscosity is $\eta = 378$ mPa·s at shear rate $\dot{\gamma} = 1$ s$^{-1}$. Initially, the film is formed in closed SE cell, without water evaporation. First, the film undergoes 4 stepwise transitions (stratification). Next, the cover of the SE cell has been removed, which leads to evaporation of water from the film. As a result, the film further thins and acquires a silvery white color, which corresponds to film thickness $h \approx 120$ nm (see SI Appendix, Table S1, for the correspondence between the film thickness and the interference colors in reflected white light). Note that the evaporation engenders a flux of water from the periphery to the center of the film, which gives rise to a hydrodynamic pressure difference that could be of the order of $10^5$ Pa between the Plateau border and the film center [40]. The initial thinning of the film in open cell can be explained with the action of this additional pressure. The further evolution of the film, which can be seen in the appended Movie S1, is as follows.

About one minute after the beginning of water evaporation, onset of local film thickening is observed near the film periphery (Fig. 1a). With time, this thickening advances and occupies a greater part of the film area (Fig. 1b). The different colors indicate nonuniform thickness. At that, a fine structure of stripes is observed, which indicates a stepwise variation of the film thickness. After that, the film acquires a vortex structure with a circular spot of thinner film in the middle and a series of colored stripes rotating around it (Fig. 1c). The rotation is observed to change direction from clockwise to anticlockwise, and back (see Movie S1). This is the stage of perfect vortex, which continues about five minutes.

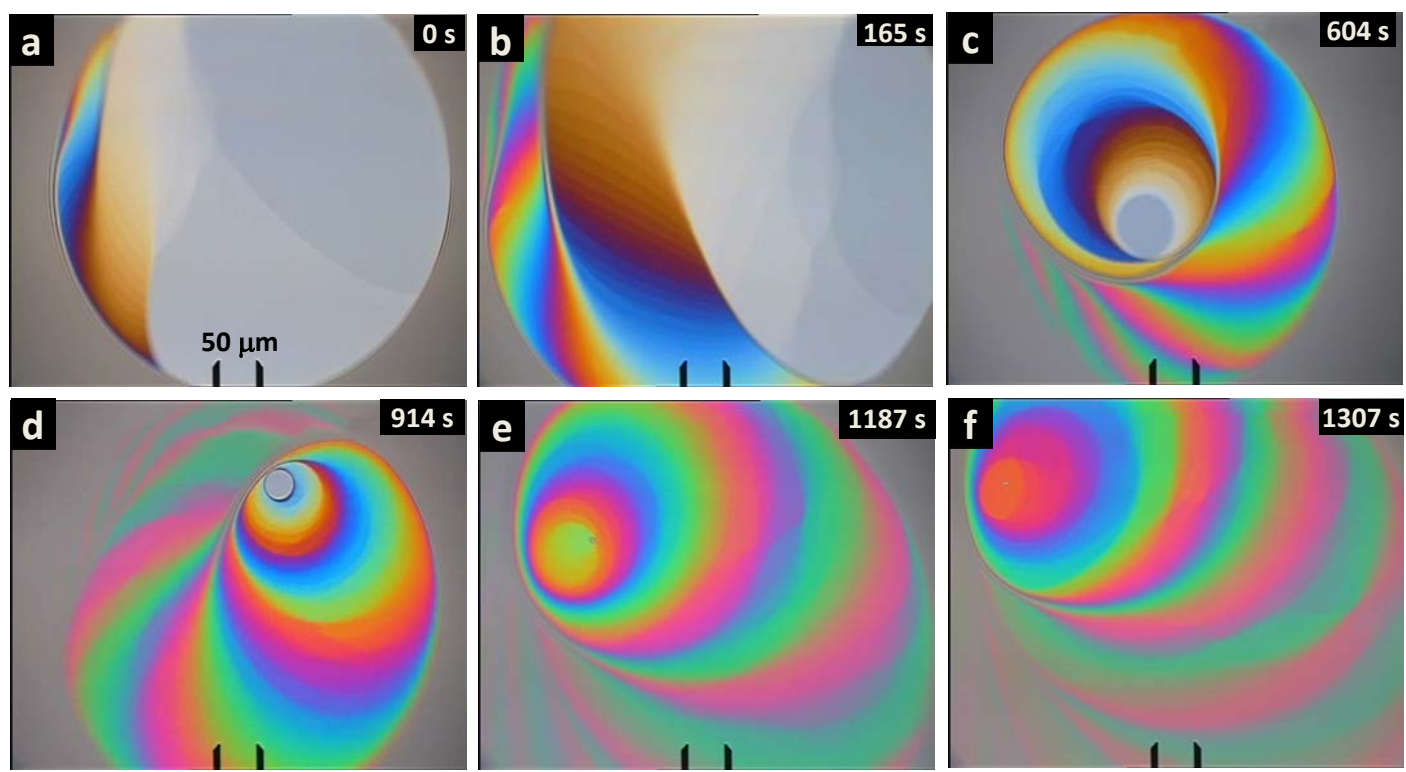

**Fig. 1.** Consecutive stages of vortex evolution in an evaporating foam film. The film is formed in SE cell from a solution of 400 mM 3:2 SLES-1EO + LB12 at pH = 6 and is observed in reflected white light. (a) Formation of thick colored film due to the penetration of liquid through the film periphery. (b) The thick film gradually occupies the whole film area and shows a fine structure of parallel stripes. (c) Stage of perfect vortex with a spot of thinner film in the center and stripes of different color (and thickness) around it. (d) With time, the area of the central spot shrinks; its silvery white color indicates film thickness of ca. 120 nm. (e) The thickness of the film increases – the yellow color of the central spot indicates local film thickness $h \approx 320$ nm. (f) Further increase of the film thickness – the orange color of the central spot indicates thickness $h \approx 350$ nm (see also Movie S1). The scaling mark is 50 μm for all photos; the upper-right marks show the development of the process in time.

Next, the spot in the center begins to shrink (Fig. 1d) and its color changes to yellow (Fig. 1e) and orange (Fig 1f), which indicates increase of the local film thickness from ca. 120 nm to 322 and 348 nm; see Fig. 2a and SI Appendix, Table S1. At that, the colors of the stripes indicate thickening of the film in the whole region of the vortex.

The micrographs in SI Appendix, Fig. S3, show that similar vortex is observed if 100 mM NaCl are added to the 400 mM 3:2 SLES-1EO + LB12 at pH = 6. Hence, the electrostatic double layer repulsion plays a secondary role in the observed phenomenon. Vortex is observed also in other studied solutions: (i) 400 mM 5:1 SLES-2EO + LB12 with added 100 mM NaCl at pH = 10.0 (SI Appendix, Fig. S4); (ii) 400 mM 5:1 SLES-2EO + LB12 at pH = 6.2 (SI Appendix, Fig. S6); (iii) 400 mM 5:1 SLES-2EO + LB12 plus added 3 wt%

DDAO at pH = 8.3 (SI Appendix, Fig. S7), and (iv) 400 mM 3:2 SLES-2EO + LB12 at pH = 6.0 (SI Appendix, Fig. S8).

For all studied systems, if we close the SE cell, the vortex disappears in the opposite order of its appearance, i.e. the phenomenon is reversible. The process ends with the formation of a plane parallel film, which has been initially observed before the opening of the experimental cell for evaporation. The vortex does not disappear immediately. This can be explained with the fact that ca. 95 % of the volume of the closed cell is air, which is saturated with water vapors for 10-15 min (at room temperature) after closing the cell. During this period of time, the vortex gradually disappears; see SI Appendix, Fig. S3 and Movie S2.

It should be also noted that in all cases the appearance of vortex is observed in films formed from slightly turbid (semi-transparent) solutions. Vortex is not observed in films formed from clear solutions (only surfactant micelles; no precipitate) and from milky-white turbid solutions (most micelles are transformed into precipitate). Hence, for the appearance of vortex it is necessary the solution to contain *both* micelles and particles.

*3.2. Characterization of meniscus profile and surfactant solutions*

As an illustration, in Fig. 2b we present a video frame from the evolution of the vortex in a foam film formed from a solution of 400 mM 5:1 SLES-2EO + LB12 plus added 3 wt% DDAO. Using the scale of colors from Ref. [38] presented in Fig. 2a (see also Table S1 in SI Appendix), one can estimate the step height of the fine structure within the vortex. The thinnest region from amber (beige) to magenta (Fig. 2b) corresponds to film thickness variation from 140 to 200 nm. The broad blue region spans thicknesses between 210 and 290 nm, followed by the yellow band 300 – 350 nm.

Note that the number of reddish bands gives the order of interference [39]. Thus, in Fig. 2a we see four reddish bands, which means that the colors in this figure belong to the first four interference orders, the last one corresponding to film thickness of ca. 780 nm. In other words, each interference order spans ca. 195 nm. For example, in Fig. 1d one could count 7 reddish bands, which means that the film thickness at its periphery is ca. 1365 nm.

In Fig. 2b, one can count 15 steps in the blue zone between the yellow (322 nm) and magenta (201 nm) stripes. Hence, the step height is ca. $(322 - 201)/15 \approx 8.1$ nm. Step height values of about 8 nm were obtained from all processed video frames of the systems with LB12 studied by us (see e.g. SI Appendix, Figs. S5 and S8).

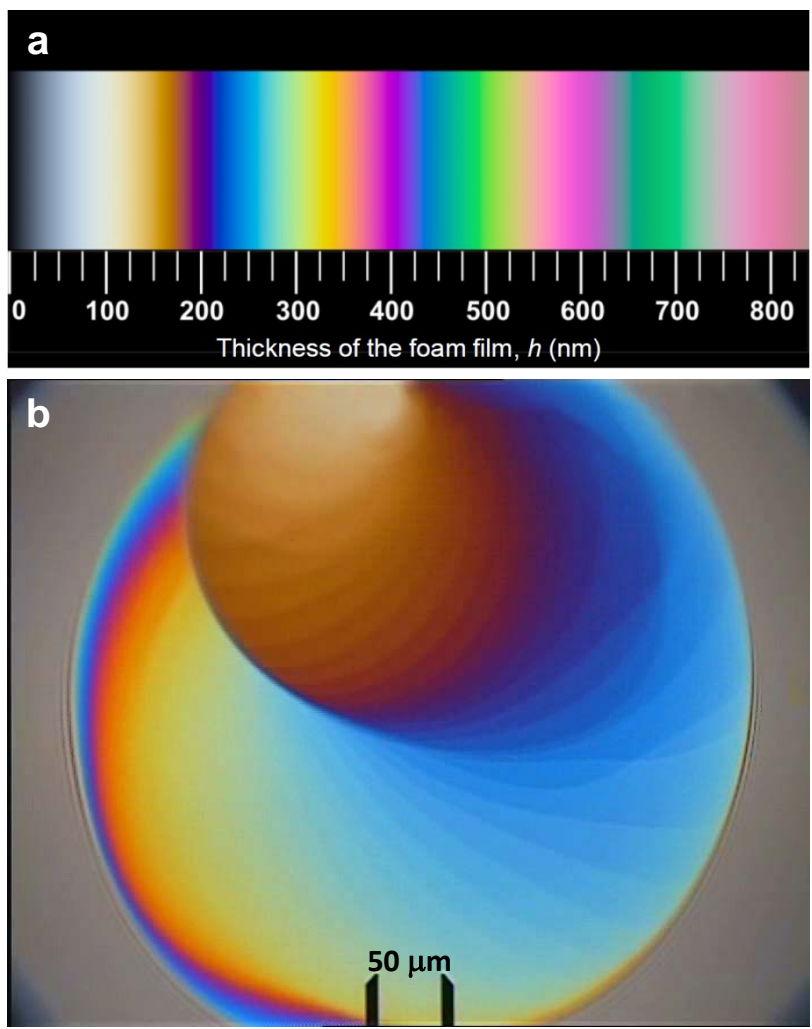

**Fig. 2.** Determination of the local film thickness using interference colors. (a) Newton series scale of two-beam interference colors in reflected white light, which has been computer generated by Zawischa [38] and is calibrated here for the refractive index of water, 1.33. The film is viewed in air at perpendicular angle of light incidence. The scale gives the approximate film thickness in nm. (b) Micrograph of the vortex in a foam film formed in SE cell from solution of 400 mM 5:1 SLES-2EO + LB12 with added 3 wt% DDAO at pH = 8.3. A fine pattern of steps is seen from the amber to the yellow zone (see SI Appendix, Table S1).

Our next goal is to see how the step height estimated from the interferograms compares with the size of the particles present in the solution. Knowing the values of the three p$K_a$ constants of LB12 (see Section 2.1) one can calculate the distribution of the four forms of this surfactant with different degrees of ionization (protonation); see Fig. 3a. One sees that the nonionic form (LB) is predominant at pH ≥ 10, whereas the trivalent cationic form ($H_3LB^{3+}$) prevails at pH ≤ 6.

Fig. 3b shows plots of the apparent viscosity $\eta$ vs. the shear rate $\dot{\gamma}$ (flow curves) measured by rotational rheometer for solutions of 400 mM 3:2 SLES-1EO + LB12 at different pH values. The natural pH of this solution is 12.3. Lower pH values have been obtained by the addition of HCl. At pH = 12.3 and 11.7, the solutions are completely transparent and the respective flow curves in Fig. 3b have a shape that is typical for surfactant solutions containing wormlike micelles [41,42]. In this pH range, the nonionic form of LB12 prevails (Fig. 3a) and the mixed micelles are negatively charged owing to SLES-1EO, which favors micellization rather than precipitation. The considerably higher viscosity at pH = 11.7 can be explained by the presence of the monovalent cation, $HLB^+$, which promotes micelle growth despite its small molar fraction at this pH. Micelle growth in catanionic mixtures is a known phenomenon [43,44].

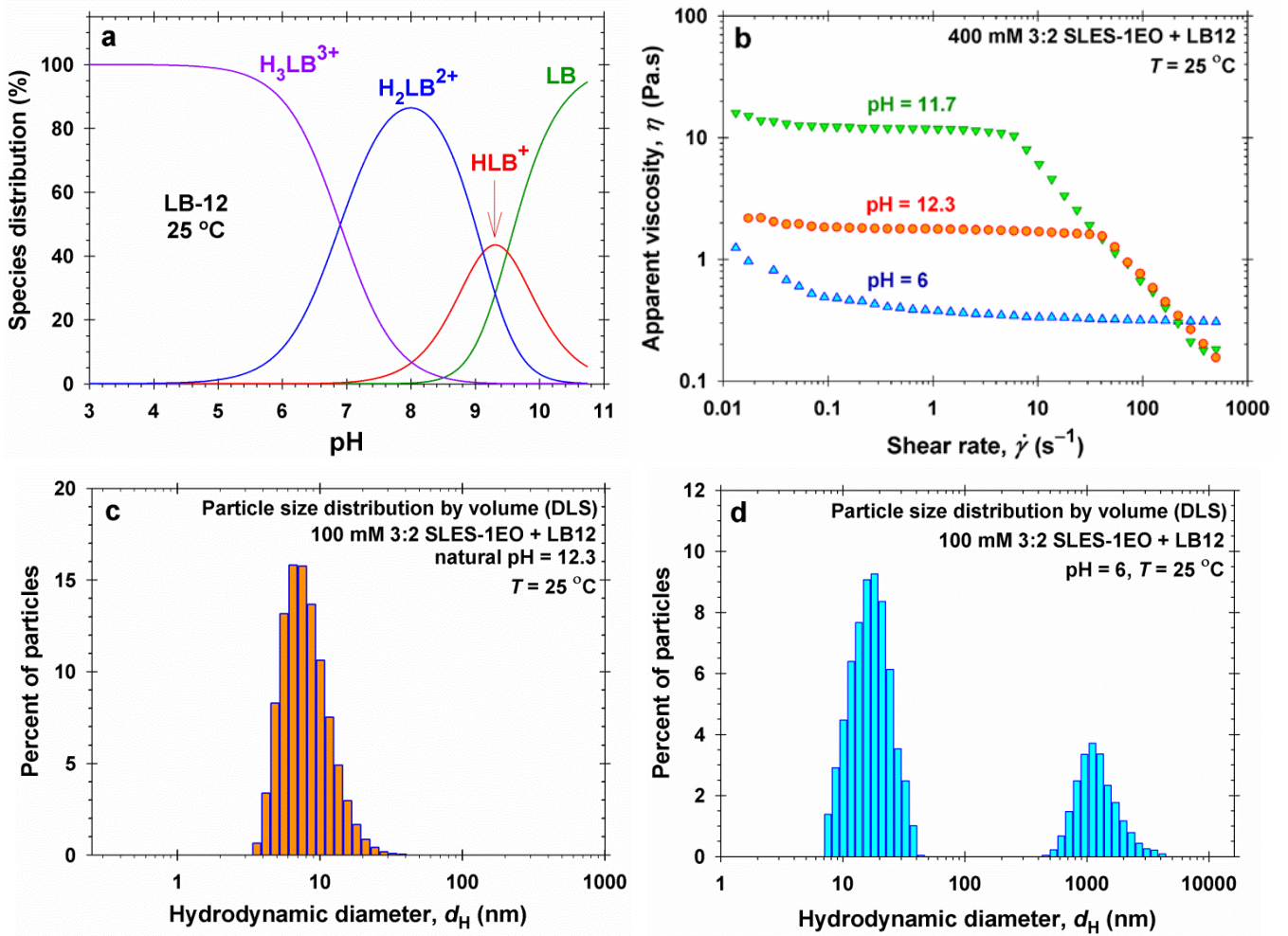

**Fig. 3.** Characterization of the used solutions. (a) Forms of LB12 of different degree of ionization (protonation): relative fraction vs. pH. (b) Flow curves: apparent viscosity $\eta$ vs. shear rate $\dot{\gamma}$ for solutions of 400 mM 3:2 SLES-1EO + LB12 at three pH values. (c) Size distribution of the particles (micelles) in a solution of 100 mM 3:2 SLES-1EO + LB12 at pH = 12.3. (d) The same, but at pH = 6.

Upon further lowering of pH, the solutions of 400 mM 3:2 SLES-1EO + LB12 become milky-white in the range 11.1 ≥ pH ≥ 7.5 (SI Appendix, Fig. S2). This can be attributed to the higher degree of ionization of the cationic LB12, which leads to enhancement of its interaction with the anionic SLES-1EO and conversion of the micelles into particles (precipitate). Vortex was not observed in this pH range.

Further lowering of pH leads to decrease of turbidity due to the domination of the positive charge of LB12 over the negative charge of SLES-1EO. This recharging, again, favors the micellization than precipitation. At pH = 6, at which vortex is observed (Fig. 1), the turbidity is rather low (SI Appendix, Fig. S2) and the viscosity is also relatively low (Fig. 3b). At pH = 6, the surface electric charge favors the formation of small, spherical or spheroidal micelles.

By dynamic light scattering (DLS), we measured the hydrodynamic diameter, $d_H$, of the particles present in the investigated solutions. Figs. 3c and 3d show the obtained particle size distributions in aqueous solutions of 100 mM 3:2 SLES-1EO + LB12 at two pH values. In comparison with Figs. 1 and 3b, these solutions have been 4 times diluted with water to ensure reliable values of $d_H$ by DLS. At pH = 12.3, the particle size distribution has a single peak at $d_H \approx 8$ nm (Fig. 3c), which is due to the mixed micelles of SLES-1EO and LB12. The position of this peak evidences that the four fold dilution has led to transformation of the elongated micelles (at pH = 12.3, Fig. 3b) into smaller spherical aggregates. Note that 8 nm is also the height of the step determined from Fig. 2b and from the experiments with stratifying films in closed cell, without evaporation (SI Appendix, Fig. S5).

At pH = 6, at which vortex is observed (Fig. 1), the particle size distribution has two peaks. The peak at $d_H \approx 1.1$ μm is due to precipitated particles, which give rise to the observed turbidity, whereas the peak at $d_H \approx 18$ nm indicates the presence of spheroidal micelles. The semi-minor axis of such spheroidal micelles is close to the radius of the spherical micelles, insofar that both of them are about one extended surfactant chainlength. For this reason, the stepwise meniscus profile in the vortex with step height ≈ 8 nm (see Figs. 2b and SI Appendix, Fig. S8) can be due to the presence of multi-layered structure of spherical or spheroidal micelles, the latter – with the long axis oriented parallel to the two film surfaces.

*3.3. Vortex with other catanionic mixtures and with silica or latex particle suspensions*

The above physical picture implies that vortex can develop in any liquid film, which is formed from solutions containing smaller and bigger colloidal particles of appropriate sizes and concentrations, i.e., in the case of bimodal particle size distribution. Vortex could be observed with other catanionic surfactant mixtures, for example, alkyl quaternary ammonium chlorides or bromides and sodium alkyl ethoxy sulfates, as well as with other mechanisms of filter cake assembly. For example, we observed well pronounced vortex in foam films from mixed solutions of SLES-2EO with dodecyl- and tetradecyl-trimethylammonium bromides (DTAB and TTAB) – see SI Appendix, Figs. S9a, S9b and Movie S3.

As already mentioned, vortex is not observed in foam films formed from transparent micellar solutions that do not contain precipitated μm-sized colloidal particles. For example, such is the solution of 400 mM 5:1 SLES-2EO/LB12 at pH = 12.5. However, after adding to the latter solution 4 wt% silica particles of mean diameter $d$ = 2.7 μm, or 4 wt% latex particles of mean diameter $d$ = 1 μm, we observed the development of vortex in the evaporating foam films – see SI Appendix, Figs. S9c and S9d. The interference patterns in the latter two figures look more disordered as compared to those obtained with vortexes in catanionic surfactant solutions without particles. The deviations from circular symmetry can be explained with solidification of the filtration cake from silica or latex particles around the film. The colors of the layers near the central spot indicate film thickness much smaller than the diameter of the μm-sized silica or latex particles, so that these layers certainly represent film stratification due to surfactant micelles.

Despite the irregularities of the pattern observed with silica or latex particles, the fact that vortex exists also in this case confirms the generality of the observed phenomenon. Its physical interpretation is subject of the rest of this article.

4. Discussion

Our experiments showed that vortex is present only if the surfactant solution is slightly turbid, i.e. if it contains both micelles and particles, the latter having mean diameter of the order of 1 μm. All colors and step-heights in the vortex (see Figs. 1 and 2, and SI Appendix, Table S1) correspond to film thicknesses smaller than the diameter of the bigger, μm-sized particles. Hence, the film contains only layers of micelles, whereas the bigger particles are located in the Plateau border (Figs. 4a and b). Their role is to support the formation of a thick film, in which the vortex develops.

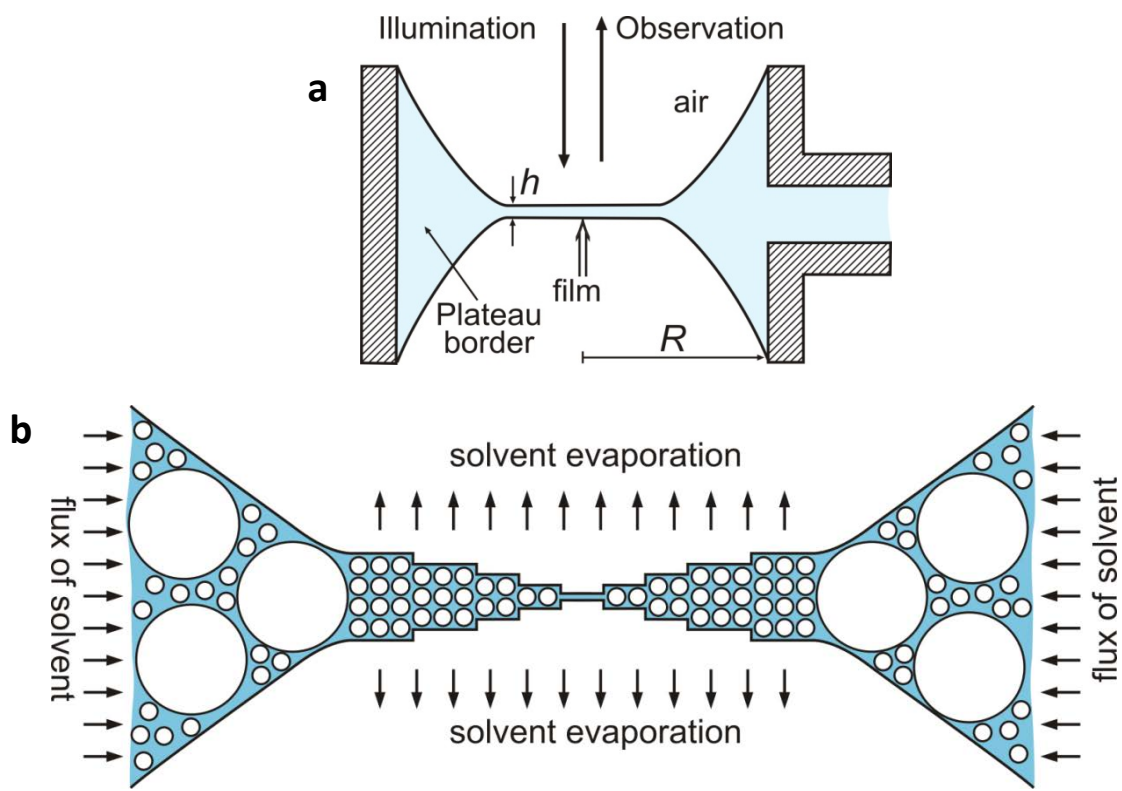

**Fig. 4.** Cross-section of the experimental cell and of a film with vortex. (a) The solutions are loaded in a cylindrical Scheludko-Exerowa (SE) cell (shown as vertical cross-section), where a liquid film is formed in the middle by sucking out the solution from the side capillary (to the right). (b) The evaporation of solvent (water) from the film gives rise to influx of water from the Plateau border toward the film. Only the small particles (the micelles) can enter the film. Their ordering, which is induced by the film surfaces, leads to stepwise film profile. The bigger (μm-sized) particles form a filtration cake around the film. The attachment of the film periphery to the edge particles of this cake allows the formation of a thick film, where the vortex develops. The flux of solvent brings more micelles in the film, which leads to a gradual increase of its thickness with time, including the thickness of the central (thinner) spot.

A possible physical picture of the observed vortex phenomenon, which emerges from our analysis of the experimental data, is shown in Fig. 4b. The hydrodynamic flux toward the film, which is provoked by the evaporation and compensates the evaporated water, carries along both micelles and μm-sized particles. The latter fill the Plateau border around the film and form a filtration cake (a porous medium) therein. The film periphery is fixed to the inner rim of this cake. Driven by the hydrodynamic flow, micelles can penetrate in the film through the filtration cake. Their volume fraction in the film is relatively high and micelle structuring gives rise to the stepwise profile of the vortex wall.

As already mentioned, the rotation of the vortex changes its direction (from clockwise to counter-clockwise and *vice versa*) several times during the observations. Hence, this rotation cannot be due to the Coriolis force. The most probable explanation is that it is driven by fluctuations in the symmetry of the evaporation-driven hydrodynamic flux. They lead to fluctuations also in the thickness of the accumulated filtration cake in the Plateau border that amplify the asymmetry of the hydrodynamic flux.

In the images shown in Figs. 1 and 2 (and in the SI Appendix), one often observes regions with very steep vortex wall. This is related to the fact that the vortex performs not only rotational, but also translational motion. Steep meniscus profile is observed either in regions where the vortex is pressed to the film periphery (against the filtration cake of the bigger particles), or in the rear part of the moving vortex, where its wall is pushed by the hydrodynamic flow of the concentrated micellar solution in the film; see Movies S1, S2 and S3.

To estimate the micelle volume fraction in the film, let us consider the simpler case of spherical micelles. Such are the micelles from the 400 mM 5:1 SLES-2EO + LB12 solution at pH = 10.0 (SI Appendix, Fig. S4). At this pH, the micelles are negatively charged thanks to the prevailing charge of the anionic SLES-2EO. The mean aggregation number of the mixed micelles can be estimated from the height of the stratification step, $\Delta h$ [13,14]:

$$N_{agg} = (c_s - \text{CMC})(\Delta h)^3 \tag{1}$$

Here, $c_s$ is the total surfactant concentration and CMC is the critical micellization concentration; when using Eq. (1), $c_s$ and CMC have to be expressed as number of molecules per unit volume. With $c_s = 400$ mM $= 0.241$ nm$^{-3}$ and $\Delta h = 8$ nm, Eq. (1) yields $N_{agg} = 123$. (For comparison, for 50 mM CTAB the aggregation number is $N_{agg} \approx 135$ [13,14].)

Eq. (1) follows from the empirical law $\Delta h = c_{mic}^{-1/3}$, where $c_{mic} = (c_s - \text{CMC})/N_{agg}$ is the concentration of surfactant micelles [7,13,14]. (In general, $\Delta h = c_p^{-1/3}$, where $c_p$ is the concentration of any charged colloidal particles, including surfactant micelles.) Then, the effective micelle volume fraction in the stratifying film is:

$$\varphi = V_{mic} c_{mic} = \frac{4}{3}\pi R^3 / (\Delta h)^3 \tag{2}$$

where $V_{mic}$ is the volume of the micelle and $R$ is its effective radius. With $R = 4$ nm and $\Delta h = 8$ nm, Eq. (2) yields $\varphi = \pi/6 \approx 0.52$.

## 5. Mechanical model for the stepwise vortex profile

*5.1. Balance of hydrodynamic and surface-tension forces*

Here, our goal is to identify the forces that determine the stepwise vortex profile; to calculate theoretically this profile, and to investigate the role of several factors that control the observed phenomenon.

All experiments in the present study are carried out at high surfactant concentrations (400 mM), much above the CMC. Under such conditions, the surface tension $\sigma$ is constant and equal to its value above the CMC. The fast kinetics of surfactant exchange between the micelles and the film surfaces suppresses any surface tension gradients [45]. For this reason, we assume that Marangoni effects driven by concentration and temperature gradients are negligible in the considered case.

The oscillatory-structural surface force is acting in direction *perpendicular* to the film surfaces. It is due to the micelle ordering in the film and determines the *height* of the stratification steps, $\Delta h$ [7–16]. The latter is equal to the effective diameter of the micelles ($\Delta h \approx 2R$).

The hydrodynamic friction force is due to the influx of water, which compensates the water evaporated from the two film surfaces. This force is acting in *radial* direction, toward the center of vortex. The hydrodynamic force would close the vortex if it was not counterbalanced by the net surface tension force, which is acting in the opposite direction, from the center of vortex toward its periphery. Here, we bring this force balance in quantitative form and demonstrate how it determines the *width* of the stratification steps.

To simplify the problem, we consider a vortex of rotational symmetry, close to that shown in SI Appendix Fig. S4. Polar coordinates $(r,\varphi)$ will be used. At this first stage of theoretical modeling, effects like vortex asymmetry and non-stationary behavior (rotational and translational motion) are not considered. The mechanical equilibrium of each ring of uniform thickness (containing a given number of micellar layers) requires the net force acting on it to be equal to zero. Following the classical Stevin force-balance approach, we can treat the respective ring as being "solidified" and to balance the external forces acting on it. At that, the net force due to surface tension, which is acting in the positive direction of the $r$-axis, is counterbalanced by the difference in the hydrodynamic pressures acting on the outer and inner boundaries of the ring (Fig. 5).

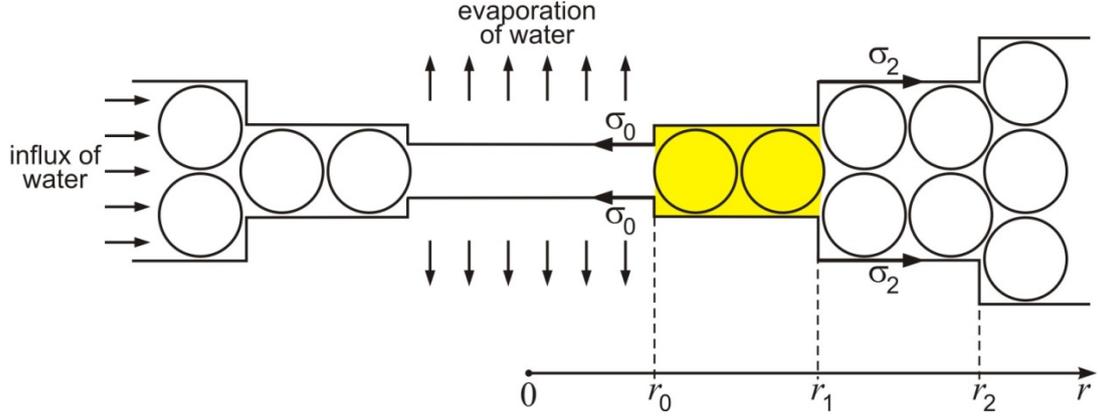

**Fig. 5.** Balance of the forces that determine the vortex profile. The ring containing one layer of micelles (see the shadowed domain) is considered as being "solidified" and the external forces acting on it are balanced; $\sigma_2$ is the surface tension of the film with two layers of micelles, and $\sigma_0$ is the surface tension of the film without micelles (details in the text).

The balance of the external forces acting in radial direction on the ring of thickness $h_1$ and width $r_1 - r_0$, which contains one layer of micelles (the shadowed domain in Fig. 5) reads:

$$2(\sigma_2 - \frac{\kappa}{r_1})r_1 d\varphi - 2\sigma_0 r_0 d\varphi = p(r_1)h_1 r_1 d\varphi - [p(r_0)h_0 + p_g(h_1 - h_0)]r_0 d\varphi \qquad (3)$$

Eq. (3) expresses the force balance on a part of an elementary sector of central angle $d\varphi$, which is situated at $r_0 \leq r \leq r_1$; $h_n$ ($n = 0, 1$) is the thickness of a film containing $n$ layers of particles; $\sigma_n = \sigma(h_n)$ is the surface tension of a film of thickness $h_n$; the outer radius of such film is denoted $r_n$ (Fig. 5); $p_g$ is the pressure in the outer gas phase; $p(r)$ is the hydrodynamic pressure acting in direction parallel to the $r$-axis. The two terms in the left-hand side of Eq. (3) represent the surface tension forces acting on the outer and inner boundaries of the considered sector, at $r = r_1$ and $r_0$; $\kappa$ is the line tension acting on the outer border, at $r = r_1$ [46]. The two terms in the right-hand side of Eq. (3) express the forces due to the hydrodynamic pressure acting on the right ($r = r_1$) and left ($r = r_0$) sides of the shadowed volume in Fig. 5.

The film surface tension can be expressed in the form [47]:

$$\sigma_n = \sigma + \frac{1}{2}\int_{h_n}^{\infty} \Pi(h)\,dh \qquad (4)$$

where $\Pi(h)$ is the disjoining pressure. As a rule, the integral in the right-hand side of Eq. (4) is negative owing to prevailing long-range attractive forces. For example, if we take into account the van der Waals forces, Eq. (4) acquires the form [9]:

$$\sigma_n = \sigma - \frac{A_H}{24\pi h_n^2} \tag{5}$$

For aqueous foam films, the Hamaker constant is $A_H \approx 3.7 \times 10^{-20}$ J [9]. With $h_n = 50$ nm, one obtains that the last term in Eq. (5) is equal to $2.0 \times 10^{-4}$ mN/m, i.e. it is negligible. Similarly, using data in Ref. [15] one could conclude that the contribution of the oscillatory-structural force to $\sigma_n$ is also negligible. For this reason, we can work with constant surface tension:

$$\sigma_n \approx \sigma, \ n = 0, 1, 2, ... \tag{6}$$

The line tension $\kappa$ can be estimated as the excess surface free energy per unit length of the contact line due to the band of width $\Delta h = h_1 - h_0$ (the height of the step):

$$\kappa \approx \sigma \Delta h = 2\sigma R \tag{7}$$

where, as usual, $R$ is the effective micelle radius. With $\sigma = 30$ mN/m and $R = 4$ nm, one obtains $\kappa = 1.2 \times 10^{-10}$ N. This value has to be compared with $(r_1 - r_0)\sigma$. Experimentally, the width of the stratification steps is of the order of micrometers, i.e. $(r_1 - r_0) \gg \Delta h$. Then, we obtain:

$$\kappa \ll (r_1 - r_0)\sigma \tag{8}$$

In other words, the term with the line tension $\kappa$ in Eq. (3) is also negligible. In view of Eqs. (6) and (8), Eq. (3) acquires the following simpler form:

$$2(r_1 - r_0)\sigma = p(r_1)h_1 r_1 - p(r_0)h_0 r_0 - 2p_g R r_0 \tag{9}$$

In a similar way, for the film of thickness $h_n$ one obtains:

$$2(r_n - r_{n-1})\sigma = p(r_n)h_n r_n - p(r_{n-1})h_{n-1} r_{n-1} - 2p_g R r_{n-1}, \ n = 1, 2, 3, ... N. \tag{10}$$

Eq. (10) represents a system of $N$ equations for determining $N$ unknowns, $r_1, r_2, ..., r_N$.

To find the pressure $p(r_n)$, we can use the Darcy's law for the liquid flow through a porous medium – the array of ordered micelles within the film [48]:

$$q(r) = K\frac{dp}{dr} \tag{11}$$

Here, $q$ (m/s) is the flow rate and $K$ is a constant. For a steady-state film, the water flux has to compensate the water evaporated from the two film surfaces; the total flux of evaporated water from a disk (film) of radius $r$ is:

$$Q(r) = 2\pi r^2 j \tag{12}$$

where $j$ is the water evaporation flux per unit area. (For example, from the data in Ref. [49] for the evaporation of water from a film of latex suspension, one can estimate $j = 180$ nm/s.) Then, the radial influx per unit time and unit area of the film cross-section is:

$$q(r) = \frac{Q}{2\pi r h_n} = \frac{r}{h_n} j \tag{13}$$

for a film of local thickness $h_n$. The combination of Eqs. (11) and (13) yields:

$$\frac{dp}{dr} = \frac{j}{K h_n} r \tag{14}$$

Integrating the above equation for a ring-shaped film of thickness $h_n$, we obtain:

$$p(r_n) - p(r_{n-1}) = \frac{j}{2 K h_n}(r_n^2 - r_{n-1}^2), \; n \geq 1 \tag{15}$$

Eq. (15) is to be combined with Eq. (10).

For a porous medium that represents a volume filled with spherical particles (micelles), the Darcy's coefficient can be calculated from the expression [48]

$$K = \frac{R^2}{45\eta} \frac{(1-\varphi)^3}{\varphi^2} \tag{16}$$

where $\eta = 0.89$ mPa.s is the viscosity of water and $\varphi$ is the volume fraction of particles (micelles). For $R = 4$ nm and $\varphi = 0.5$, Eq. (16) yields $K = 2.0 \times 10^{-16}$ m$^2$/(Pa.s).

*5.2. Solution to the problem and numerical results*

To obtain numerical results, it is convenient to introduce dimensionless variables:

$$\tilde{r}_n = \frac{r_n}{r_0}, \; \tilde{h}_n = \frac{h_n}{h_0}, \; \text{and} \; \tilde{p}_n = \frac{p_n}{p_g} \; (n \geq 0) \tag{17}$$

The dimensionless form of Eq. (15) is:

$$\tilde{p}_n - \tilde{p}_{n-1} = J \frac{\tilde{r}_n^2 - \tilde{r}_{n-1}^2}{\tilde{h}_n}, \; n \geq 1 \tag{18}$$

where $J$ is a dimensionless evaporation flux:

$$J \equiv \frac{r_0^2 j}{2 K p_g h_0} \tag{19}$$

Likewise, Eq. (10) acquires the form:

$$(\tilde{r}_n - \tilde{r}_{n-1})S = \tilde{p}_n \tilde{h}_n \tilde{r}_n - \tilde{p}_{n-1} \tilde{h}_{n-1} \tilde{r}_{n-1} - H \tilde{r}_{n-1} \tag{20}$$

where $S$ is dimensionless surface tension and $H$ is dimensionless step height:

$$S \equiv \frac{2\sigma}{p_g h_0} \quad \text{and} \quad H \equiv \frac{2R}{h_0} \tag{21}$$

Next, it is convenient to represent Eq. (20) in the following equivalent form:

$$(\tilde{p}_n - \tilde{p}_{n-1}) \tilde{h}_n \tilde{r}_n + [\tilde{p}_{n-1}(\tilde{h}_{n-1} + H) - S](\tilde{r}_n - \tilde{r}_{n-1}) = (1 - \tilde{p}_{n-1}) H \tilde{r}_{n-1} \tag{22}$$

Finally, the substitution of Eq. (18) in Eq. (22) yields:

$$[J \tilde{r}_{n-1}^2 (x_n + 1) x_n + \tilde{p}_{n-1}(\tilde{h}_{n-1} + H) - S](x_n - 1) = (1 - \tilde{p}_{n-1}) H, \quad n \geq 1 \tag{23}$$

$$x_n \equiv \frac{\tilde{r}_n}{\tilde{r}_{n-1}} > 1 \tag{24}$$

Eq. (23) is a cubic equation for $x_n$, which allows one to determine the vortex profile by means of a recurrence procedure at given $J$, $H$ and $S$. We start with given values of $r_0$, $h_0$ and $p_0$ for the thinnest central film ($p_0 < p_g$). In view of Eq. (17), we obtain:

$$\tilde{r}_0 = 1, \quad \tilde{h}_0 = 1, \quad \text{and} \quad \tilde{p}_0 < 1 \tag{25}$$

Furthermore, Eq. (23) is solved and $x_1$ is determined. Next, $\tilde{r}_1 = x_1 \tilde{r}_0$; $\tilde{h}_1 = \tilde{h}_0 + H$, and $\tilde{p}_1$ is determined from Eq. (18). Furthermore, we solve Eq. (23) for $n = 2$ to determine $x_2$, and so on and so forth.

In Eq. (23), $(x_n - 1) > 0$ and $(1 - \tilde{p}_{n-1}) > 0$. Hence, in order to have solution for $x_n$, the term in the brackets must be also positive. For $n = 1$, this requirement is fulfilled if

$$J \tilde{r}_0^2 (x_1 + 1) x_1 > S \approx S - \tilde{p}_0 (1 + H) \tag{26}$$

Here, we have used the fact that $S \gg \tilde{p}_0(1 + H)$ for typical parameter values. In view of Eq. (24), we have $x_1 > 1$, so that $J(x_1 + 1)x_1 > 2J$. Hence, Eq. (26) will be fulfilled if

$$2J > S \;\Rightarrow\; r_0^2 > \frac{2K\sigma}{j} \tag{27}$$

At the last step, we used the expressions for $J$ and $S$ in Eqs. (19) and (21). Eq. (27) leads to the following conclusions:

(i) The radius of the central spot of the vortex, $r_0$, cannot be smaller than $(2K\sigma/j)^{1/2}$; otherwise, the hydrodynamic and surface tension forces cannot be balanced.

(ii) The lower the evaporation flux, $j$, the greater should be $r_0$. In other words, if $j$ decreases (say, when approaching the state of equilibrated water vapors), then the central spot should expand driven by the surface tension force, because the counteracting hydrodynamic force has weakened. This is observed experimentally – see SI Appendix, Fig. S3 (closed cell).

(iii) The thickness of the central spot, $h_0$, does not enter the limitation for $r_0$, Eq. (27). This result correlates with the fact that in many cases the central spot corresponds to thickness above 100 nm. For example, in SI Appendix Fig. S9 the central spot has blue color, which corresponds to $h_0 \approx 250$ nm. Note, however, that $h_0$ affects the steepness of the vortex profile – see Fig. 6a.

Fig. 6 illustrates the effect of $h_0$, $p_0$ and $j$ on the vortex profiles predicted by Eq. (23). In the model calculations, the following parameter values have been used: pressure in the gas phase $p_g = 1$ atm $= 101325$ Pa; surface tension $\sigma = 30$ mN/m; $r_0 = 30$ µm; step height $\Delta h = h_n - h_{n-1} = 8$ nm ($n \geq 1$), and Darsy's coefficient $K = 2.0 \times 10^{-16}$ m$^2$/(Pa.s) – see Eq. (16).

Physically, $h_0$ is determined from the normal force balance, whereas $r_0$ – from the tangential force balance. The values of $r_0$ and $j$ obey Eq. (27). In our model calculations, $r_0$ has been fixed to a typical experimental value, 30 µm, whereas $h_0$, $p_0$ and $j$ have been independently varied to investigate their effect on the stepwise vortex profile; the used values are shown in the graphs (Fig. 6). The values of $h_0$ are in the range of spot thicknesses determined from the experimental interference patterns.

Fig. 6a indicates that if the thickness of the central spot, $h_0$, is smaller, the vortex wall is steeper. This is observed experimentally – compare the vortex profiles with smaller and larger central spot in SI Appendix Fig. S4. Note that $p_0$ could be negative. In general, the pressure inside a thin liquid film is anisotropic (tensorial), i.e. the pressures acting in directions parallel and normal to the film surfaces are different. The tangential pressure $p_0$ is lower than the normal pressure, $p_g$, and their difference equals the disjoining pressure, $\Pi = p_g - p_0$; see e.g. Refs. [50,51].

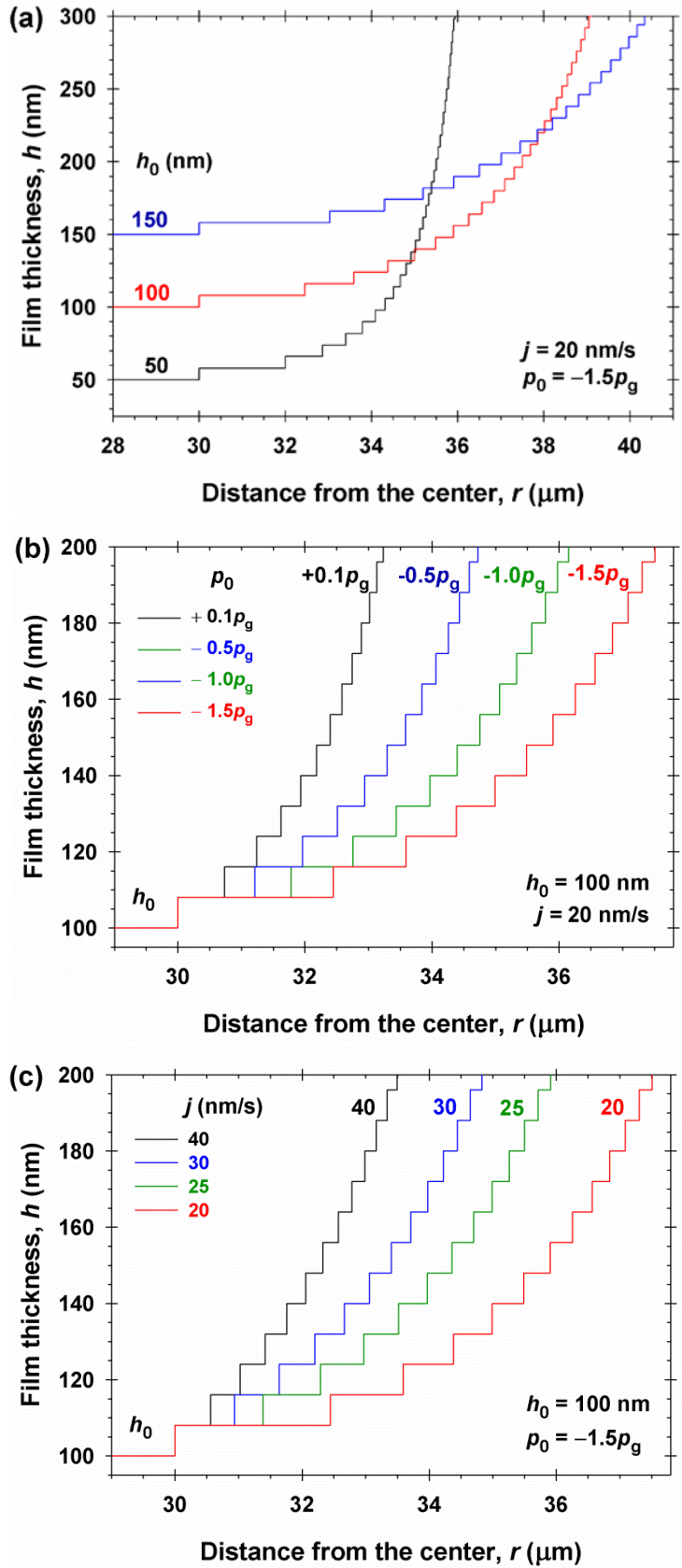

**Fig. 6**. Calculated vortex profiles, $h$ vs. $r$. (a) Effect of the film thickness in the central spot, $h_0$. (b) Effect of the pressure inside the film of the central spot, $p_0$. (c) Effect of the rate of water evaporation characterized by the evaporation flux, $j$. The values of the fixed parameters are shown in the right-down corner of each graph; details in the text.

Fig. 6b illustrates the effect of the pressure in the central spot, $p_0$: higher $p_0$ corresponds to steeper vortex profile. Fig. 6c illustrates the effect of the evaporation rate, characterized by the evaporation flux $j$. For example, $j = 20$ nm/s means that during each second an aqueous layer of thickness 20 nm is evaporated from the film surface. Because we consider stationary regime, the evaporated water is supposed to be completely compensated by the influx of water from the Plateau border. Greater $j$ leads to steeper vortex profile (Fig. 6c).

Finally, we recall that our model considers the simplest case of axisymmetric and stationary vortex. As demonstrated here, even this simplified model gives quantitative physical information for the basic acting forces and for the effect of different parameters on the vortex profile.

## 6. Conclusions

The existence of a new phenomenon – vortex in evaporating free liquid films is reported. The films are formed from solutions containing both nanosized surfactant micelles and micronsized colloidal particles, which appear in mixed solutions of cationic and anionic surfactants, where micelles coexist with precipitated particles.

The driving force of the observed phenomenon is the water evaporation from the film surfaces. The bigger particles form a filtration cake in the Plateau border, which supports the thick film ($h > 300$ nm), where the vortex develops. The ordering of the smaller particles (the micelles) in this film leads to a stepwise profile of the vortex wall with up to 20–30 stratification steps of step-height equal to the effective micellar diameter (ca. 8 nm). For thickness greater than 100 nm, stratification in foam films from micellar solutions has never been observed in previous studies [1,2,13,14,16,52,53].

Vortex was observed not only in films from catanionic surfactant solutions (SI Appendix, Figs. S3, S4, S6-S8, S9a,b) but also in films from silica and latex particle suspensions, which contain also smaller surfactant micelles (Figs S9c,d). In all studied systems, vortex appears in films from colloidal dispersions with bimodal particle size distribution – the small particles are micelles, whereas the bigger ones are colloidal particulates: surfactant precipitate or silica/latex spheres.

Theoretical model is developed for the simpler case of axisymmetric and stationary vortex (Section 5). The model demonstrates that the vortex stepwise profile, which is described by Eq. (23), is determined by the balance of hydrodynamic and surface tension forces. The effect of various factors, such as the thickness and inner pressure of the central spot, $h_0$ and $p_0$, and the evaporation rate, $j$, has been investigated; see Fig. 6.

The investigated phenomenon could find applications in the following three directions:

(i) Method for visualization and investigation of particle/micelle structuring in liquid films: The information from the quantified stepwise vortex profile complements the information obtained by other methods – formation and expansion in spots in liquid films [1–4,13,14,16,53] and measurements of structural forces by colloidal-probe AFM [5-7,20].

(ii) Foam evolution and stabilization: The thickening of liquid films owing to the influx of micelles (Figs. 1d – 1f) leads to a natural mechanism of stabilization of foams subjected to evaporation. The formation of filtration cakes from the bigger particles in the Plateau borders also produces a foam-stabilizing effect through blocking of the liquid drainage out of the foam. The appearance of vortex can be a part of the evolution of drying foams from suspensions of colloid particles + surfactant micelles, which are used to produce heat-insulation porous materials [54].

(iii) Engineering of nanostructures with amphitheatrical architecture: By polymerization [55] or freezing [56] of the film with vortex, one could produce microstructures of rotational symmetry and nano-stepwise wall profile.

Future studies could include experimental investigations on the vortex phenomenon with other colloidal systems of bimodal particle size distribution and application of this phenomenon in the aforementioned three research directions. Theoretically, in addition to the radial flow considered in our model, the flow in circumferential direction has to be taken into account, which would lead to quantitative description of the rotational and translational motion of the vortex.

**Acknowledgements**

This work was funded by Lonza Arch UK Biocides Ltd., Blackley (Project EXHIBIT A-4). Financial support is acknowledged also from the Operational Programme "Science and Education for Smart Growth", Bulgaria, grant numbers BG05M2OP001-1.002-0012 (for P.A.K.) and BG05M2OP001-1.002-0023 (for K.D.D.). The authors are grateful to Prof. Dr. Dietrich Zawischa for providing us with the scale of interference colors of high resolution computed by him [38] and to MSc Nikola Alexandrov for his help in the electronic processing of video frames and movies.

**Supplementary Information**

Supplementary data for this article can be found in the Appendix – see below.

# Supplementary Information for the article

## Vortex in liquid films from concentrated surfactant solutions containing micelles and colloidal particles

E. S. Basheva, P. A. Kralchevsky, K. D. Danov, R. D. Stanimirova, N. Shaw, J. T. Petkov

Corresponding author: Peter A. Kralchevsky
Email: pk@lcpe.uni-sofia.bg

**The Appendix includes:**
    Supplementary text
    Figures S1 to S9
    Table S1
    Legends for Movies S1, S2 and S3
    Reference

**Other supplementary material for this manuscript includes:**
    Movies S1, S2 and S3

# APPENDIX

**Materials**

The used tri-amine cationic surfactant N-(3-aminopropyl)-N-dodecyl-1,3-propane-diamine, known as Lonzabac-12 (LB12), product of Lonza (Basel); see **Fig. S1**. It was used in mixture with the anionic surfactant sodium laurylethersulfate with one or two ethylene oxide groups (SLES-1EO or SLES-2EO), **Fig. S1**, with commercial names STEOL®CS-170 and STEOL®CS-270, products of Stepan Co., USA. The other used materials are as follows: The zwitterionic surfactant dodecyl-dimethylamine oxide (DDAO) and the cationic surfactants dodecyl and tetradecyl-ammonium bromide (DTAB and TTAB), all of them products of Sigma Aldrich. We used also silica particles Excelica UF-305 of mean diameter 2.7 $\mu$m produced by Tokuyama Co., Japan, and polystyrene latex particles with sulfate surface groups and mean diameter 1 $\mu$m, product of Dow Chem. Co.

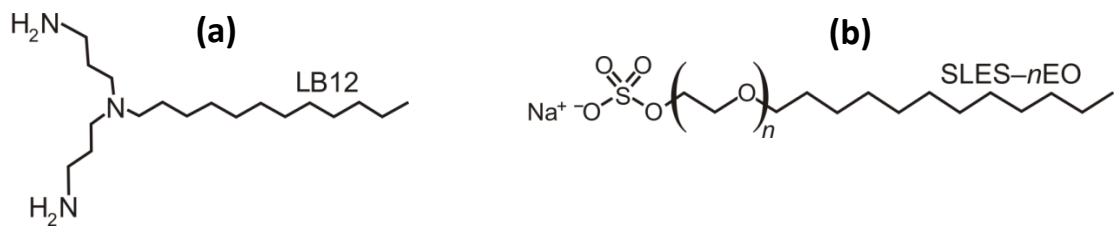

**Fig. S1.** Mixed aqueous solutions of (a) the cationic surfactant N-(3-aminopropyl)-N-dodecyl-1,3-propanediamine, Lonzabac-12 (LB12) with (b) the anionic surfactant sodium laurylethersulfate with $n$ ethylene oxide groups (SLES–$n$EO, $n$ = 1,2) have been used.

**Turbidity and viscosity of 400 mM 3:2 SLES-1EO + LB12 solutions at various pH values.** Fig. S2 illustrates the phase behavior of the studied solutions. They are transparent for 12.35 ≥ pH ≥ 11.43; turbid (milky white) for 11.1 ≥ pH ≥ 7.5 and slightly turbid for pH ≈ 6, where the vortex has been observed.

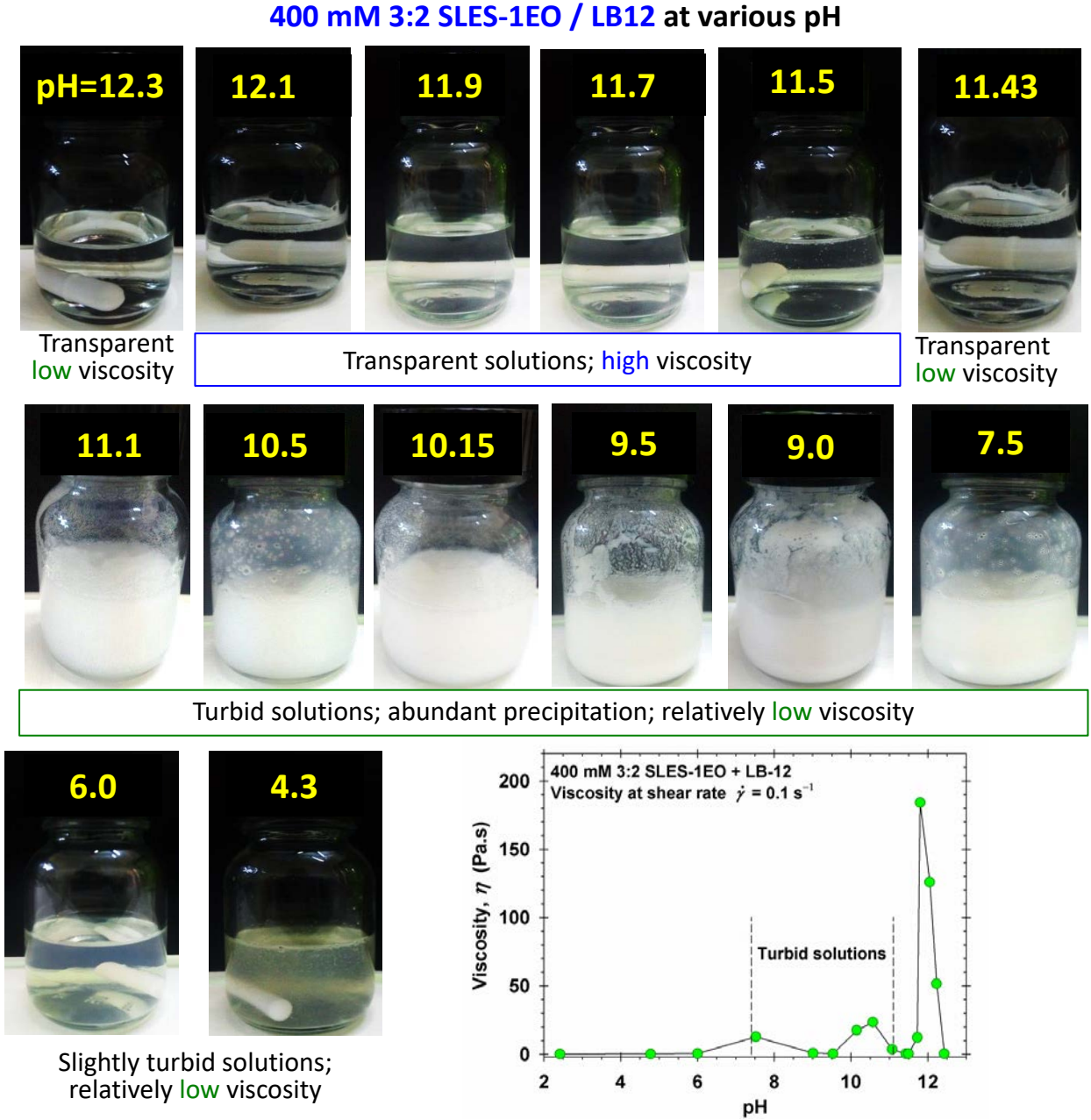

**Fig. S2. Turbidity and viscosity of 400 mM 3:2 SLES-1EO + LB12 solutions at various pH values.** The very turbid (heavily precipitated) solutions of this catanionic mixture are in the range 7.5 ≤ pH ≤ 11.1. The variation of viscosity $\eta$ with pH is shown in the graph. Vortex is observed in films formed from slightly turbid solutions of relatively low viscosity, like that at pH = 6.

**Experiments with liquid films**

**Table S1**. Interference colors and the respective thicknesses of the foam film [1].

| Color* | Film thickness $h$ (nm) | Color* | Film thickness $h$ (nm) |
|---|---|---|---|
| Black | 0 – 50 | Orange | 348 |
| Silvery white | 120 | Crimson | 371 |
| Amber | – | Purple | 396 |
| Magenta | 201 | Blue | 410 |
| Violet | 216 | Blue | 428 |
| Blue | 260 | Emerald green | 466 |
| Green | 290 | Yellow green | 502 |
| Yellow | 322 | Carmine | 542 |

*The groups of rows marked by different colors correspond to different interference orders.

*Evolution of a vortex in a film from solution of 400 mM 3:2 SLES-1EO + LB12 with added 100 mM NaCl at pH = 6*. **Fig. S3** demonstrates the reversibility of the phenomenon vortex in liquid film. The vortex appears in open experimental cell, where evaporation of water from the film takes place; see also **Fig. S4**. If the experimental cell is closed and the evaporation stops, the vortex disappears in the opposite way of its appearance.

*Stepwise thinning (stratification) of a film from a transparent solution of 400 mM 5:1 SLES-2EO + LB12 at pH = 12.4*. **Fig. S5** illustrates the experimental fact that if the solution is transparent (no precipitate), vortex is not observed. Instead, one observes stepwise thinning (stratification). The first transitions happen in closed cell. The next transitions and further film thinning are observed after opening the cell. The mean height of the step, 8.2 nm, practically coincides with the step height on the vortex profile and is close to two lengths of surfactant molecules, i.e. to the micellar diameter; the micelles are negatively charged and spherical.

*Evolution of vortex in a film from a slightly turbid solution of 400 mM 5:1 SLES-2EO + LB12 at pH = 6.2*. **Fig. S6** corresponds to the same surfactant concentrations as in Fig. S4, but now the pH is lower and the solution is slightly turbid. The appearance of vortex in this case illustrates the experimental fact that vortex appears if the solution contains not only surfactant micelles, but also bigger colloidal particles

*Evolution of vortex in a film from a slightly turbid solution of 400 mM 5:1 SLES-2EO + LB12 + added 3 wt% DDAO at pH = 8.3*. **Fig. S7** presents consecutive video frames illustrating the evolution of a typical vortex in liquid film, including the thickening of the central spot and of the whole film with time. It is seen that the vortex wall has the stepwise profile up to the yellow stripe, which corresponds to 322 nm. This corresponds to ca. 38 stratification steps. In these experiments, the greatest numbers of steps in stratifying films have been observed so far. (Usually, one observes up to 5 stepwise transitions, as in Fig. S4.)

*Estimation of the step height for a vortex in a film from a solution of 400 mM 3:2 SLES-2EO + LB12 at pH = 6*. Knowing the film thickness corresponding to the different coloured stipes (**Table S1** and Fig. 2a in the main text) and counting the number of steps between two coloured stripes, one can estimate the height of the step. As demonstrated in **Fig. S8**, the step height determined in this way turns out to be 8.3 nm, which practically coincides with the step height determined in Figs. S4 and Fig. 2b, as well as with the diameter (thickness) of the cylindrical micelles present in these solutions.

**400 mM 3:2 SLES-1EO / LB12 + 100 mM NaCl; pH = 6**

**Open cell**

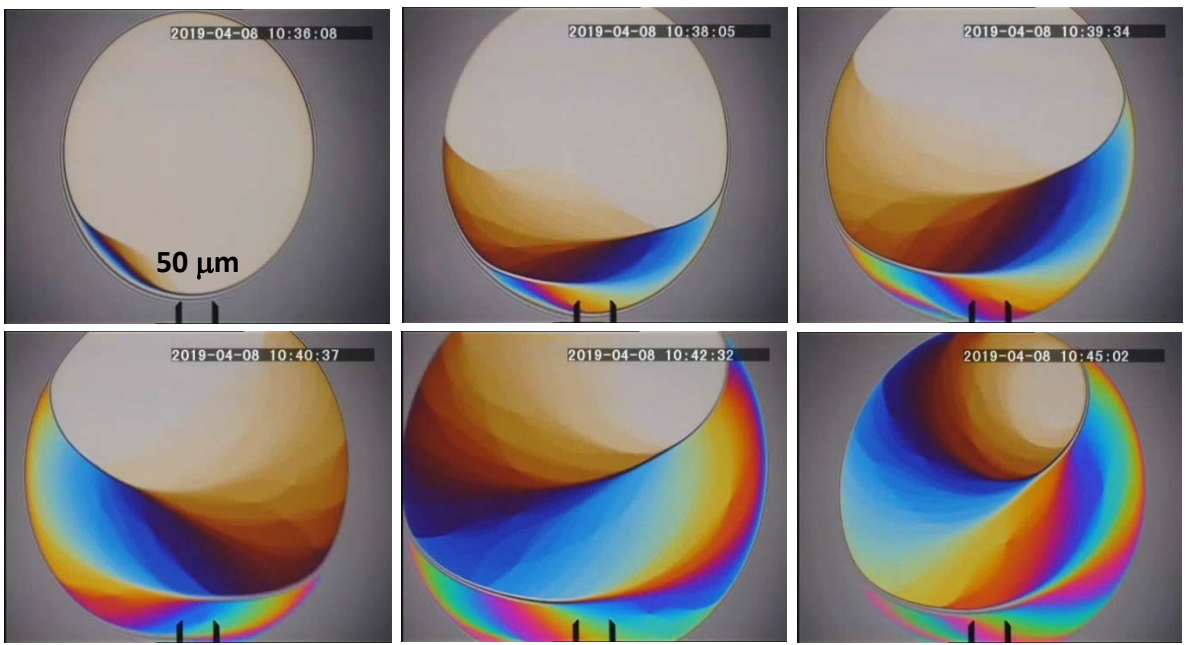

**Closed cell**

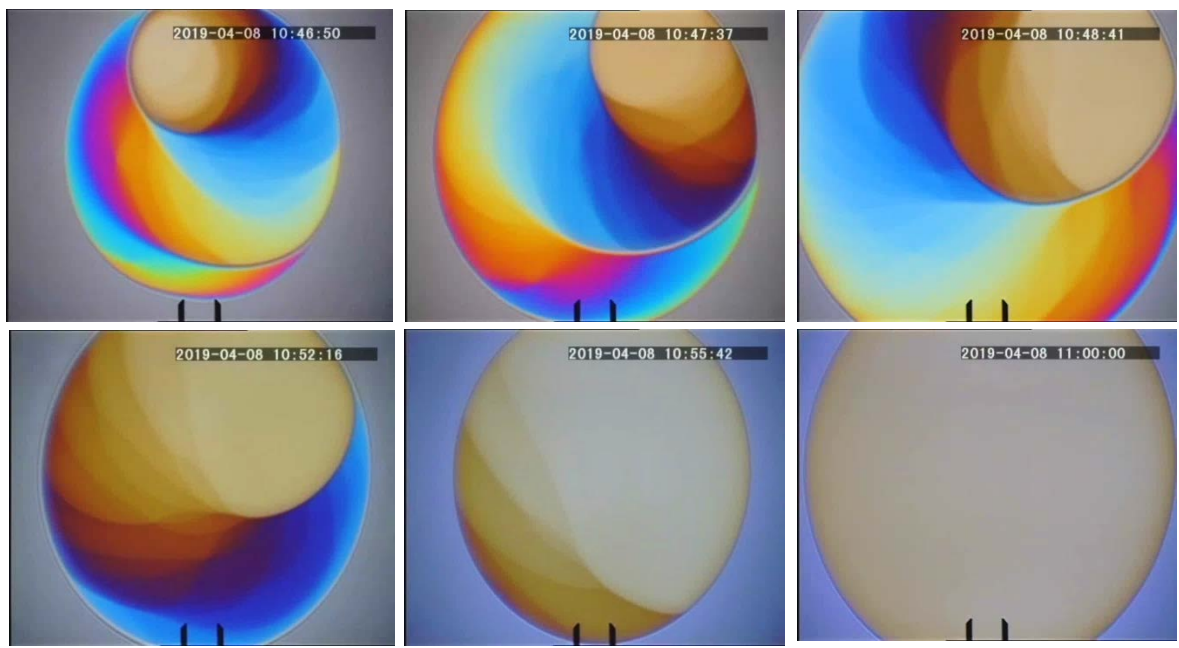

**Fig. S3. Evolution of a vortex in a film from solution of 400 mM 3:2 SLES-1EO + LB12 with added 100 mM NaCl at pH = 6.** The consecutive video frames illustrate that in open SE cell (with water evaporation) the vortex develops after thickening of an initial film of amber color and thickness ca. 160 nm. After closing the SE cell (upon gradual saturation of the water vapors in the cell and ceasing the evaporation from the film), the vortex disappears in the opposite way of its appearance (see also Movie S2). Reference distance = 50 μm.

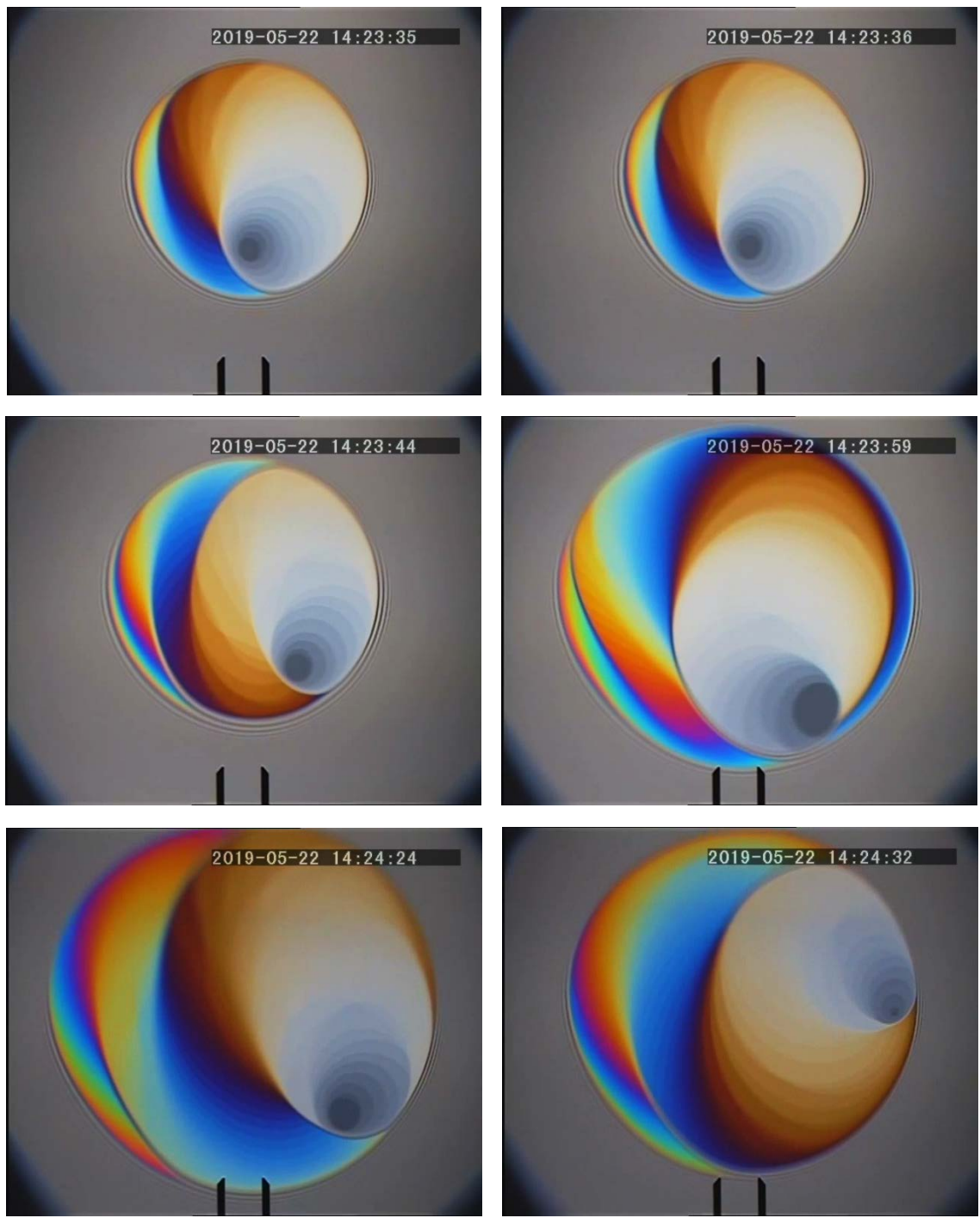

**Fig. S4. Development and evolution of a vortex in a film from solution of 400 mM 5:1 SLES-2EO + LB12 with added 100 mM NaCl at pH = 10.0 in open cell** (with water evaporation). The consecutive video frames illustrate the development of vortex with stepwise wall profile in the grey zone of thickness $50 \leq h \leq 120$ nm; the thinnest black central spot has thickness $h \leq 50$ nm. Reference distance = 50 μm.

**Film formed from transparent solution (no precipitate)**

**Viscosity $\eta$ = 0.162 Pa·s at $\dot{\gamma}$ = 0.1 s$^{-1}$**

**Closed cell**       400 mM 5:1 SLES-2EO/LB12 at pH = 12.4 (natural)

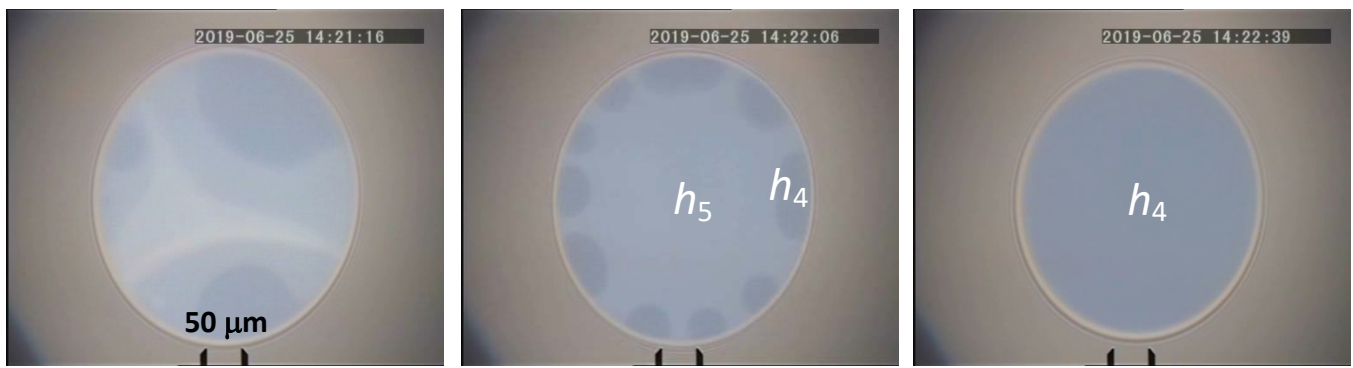

**Open cell**

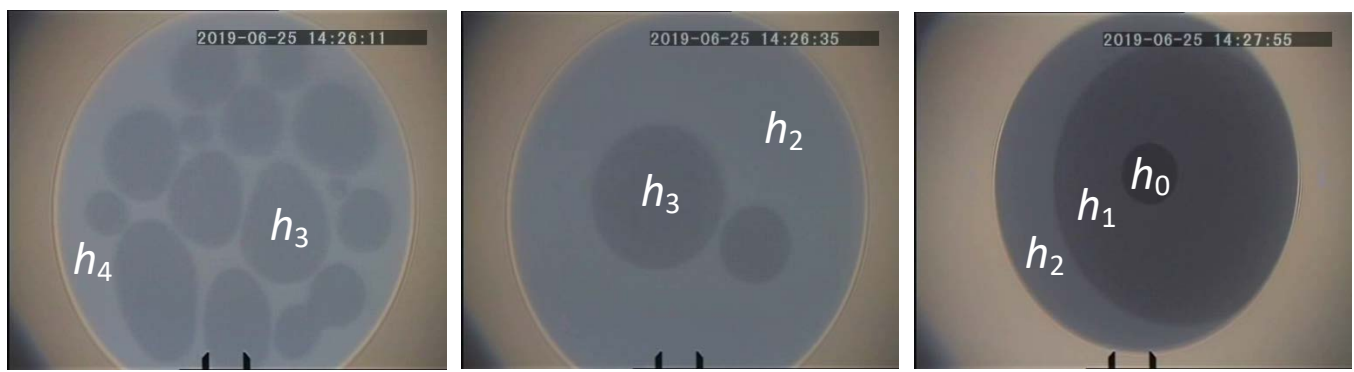

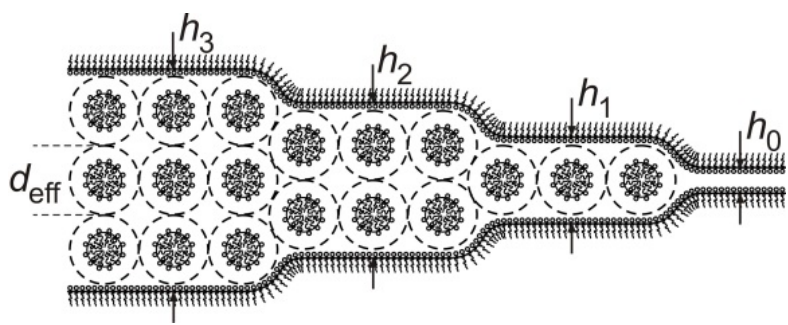

**Fig. S5. Stepwise thinning (stratification) of a film from a transparent solution of 400 mM 5:1 SLES-2EO + LB12 at pH = 12.4.** The stepwise thinning occurs as consecutive appearance and expansion of dark spots (of smaller thickness), which are due to portions of the film, which contain different number of micelles, as illustrated in the sketch. The mean height of a step is 8.2 nm. Reference distance = 50 μm.

**Film formed from 400 mM 5:1 solution of SLES-2EO/LB12 pH = 6.2;**

**slightly turbid solution; viscosity $\eta$ = 0.436 Pa·s at $\gamma$ = 0.1·s$^{-1}$**

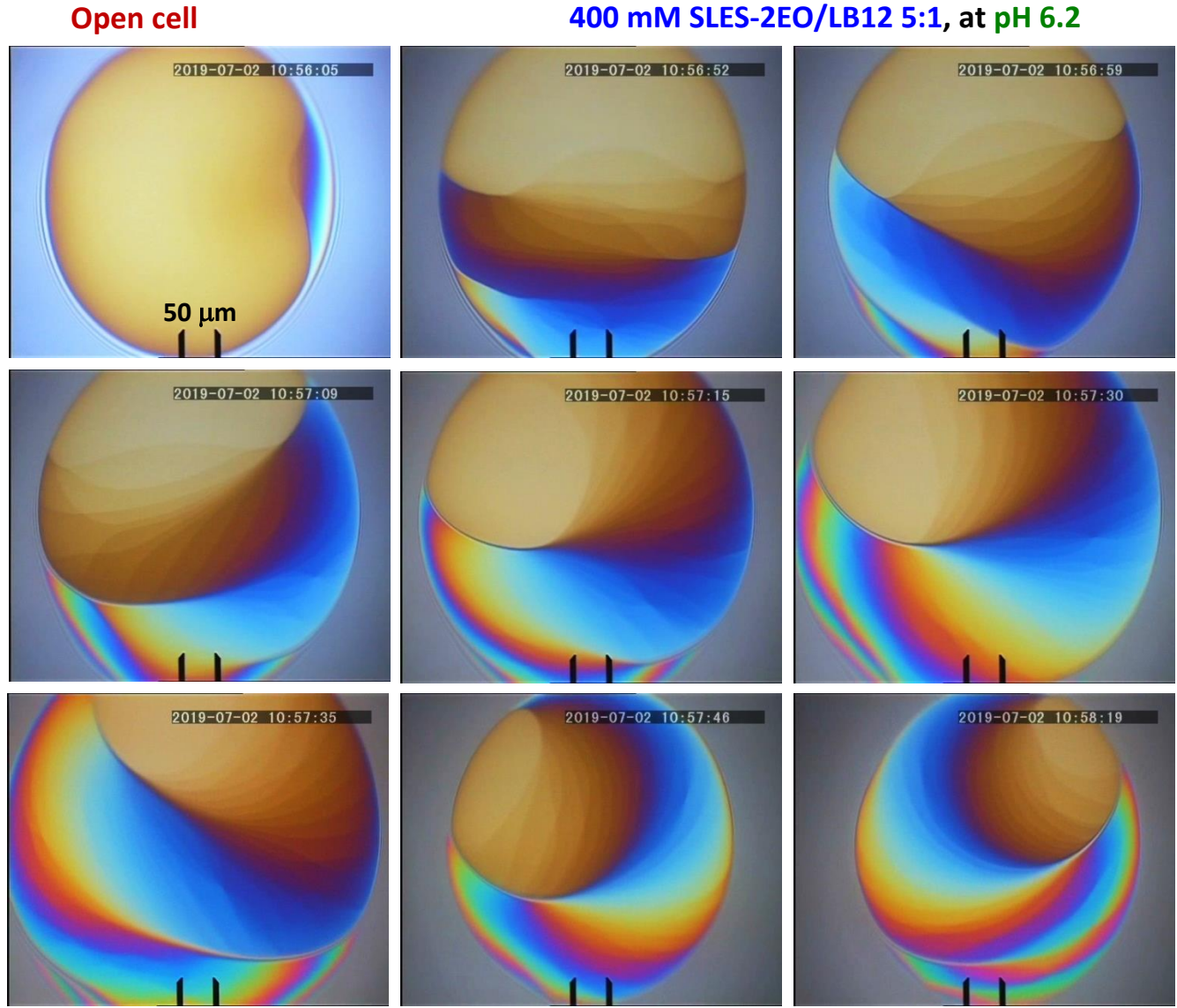

**Fig. S6. Evolution of vortex in a film from a slightly turbid solution of 400 mM 5:1 SLES-2EO + LB12 at pH = 6.2.** This experiment illustrates the fact that vortex appears if the solution contains not only surfactant micelles, but also bigger colloidal particles, which give rise to the observed slight solution's turbidity. Reference distance = 50 μm.

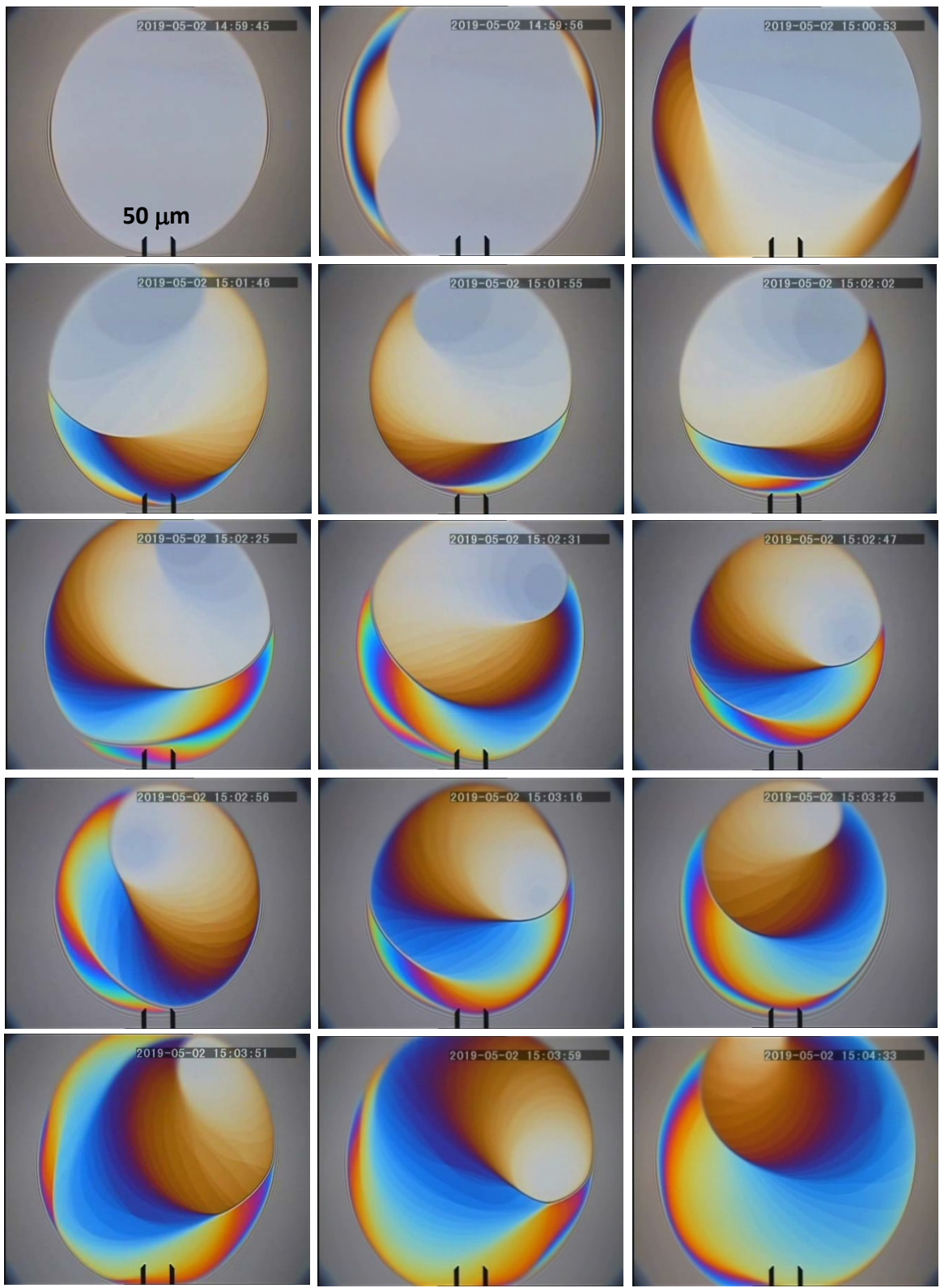

**Fig. S7. Evolution of vortex in a film from a slightly turbid solution of 400 mM 5:1 SLES-2EO + LB12 + added 3 wt% DDAO at pH = 8.3.** Reference distance = 50 μm.

**400 mM 3:2 SLES-2EO/LB12 at pH = 6**

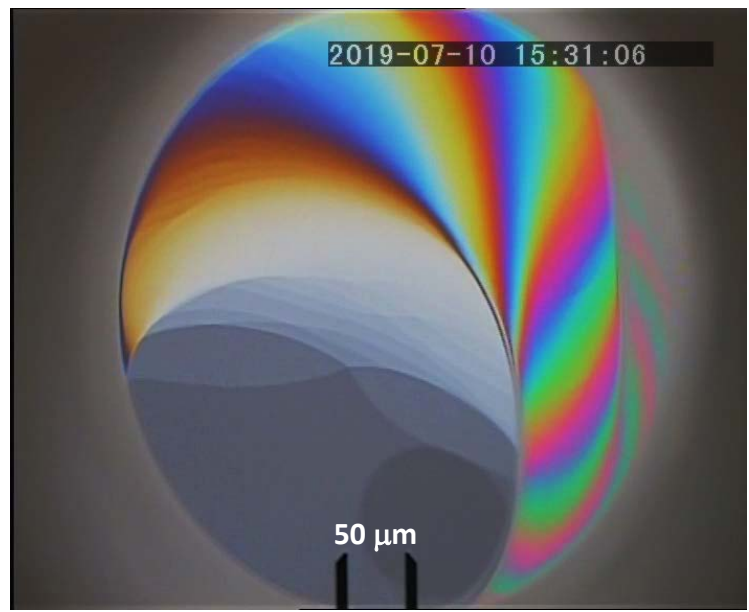

11 steps in the grey zone

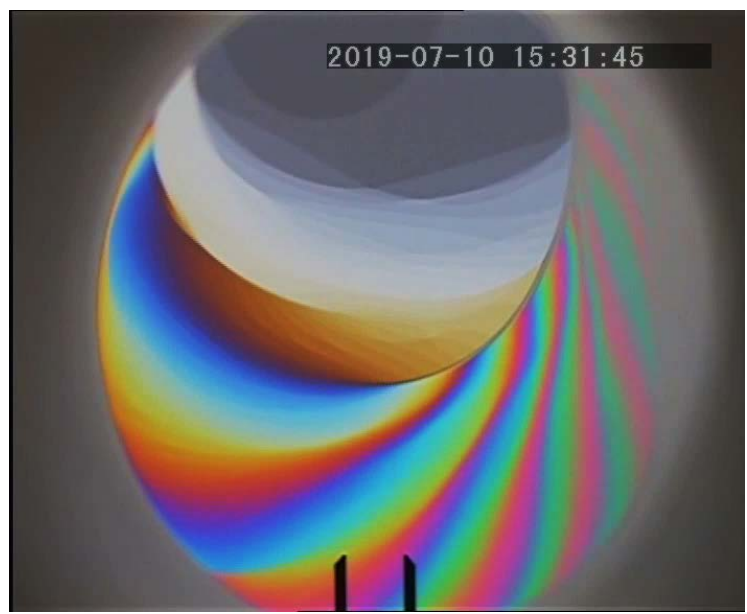

12 steps in the amber–magenta zone

**Fig. S8. Estimation of the step height for a vortex in a film from a solution of 400 mM 3:2 SLES-2EO + LB12 at pH = 6.** The number of steps is total: 11 + 12 = 23 steps. They correspond to a thickness 201 nm (magenta) – 10 nm (black film) = 191 nm. They correspond to a thickness 201 nm (magenta) – 10 nm (black film) = 191 nm. The height of a step, 191/23 = 8.3 nm, corresponds to micelle diameter determined from the stratification steps observed in closed cell; see e.g. Fig. S5.

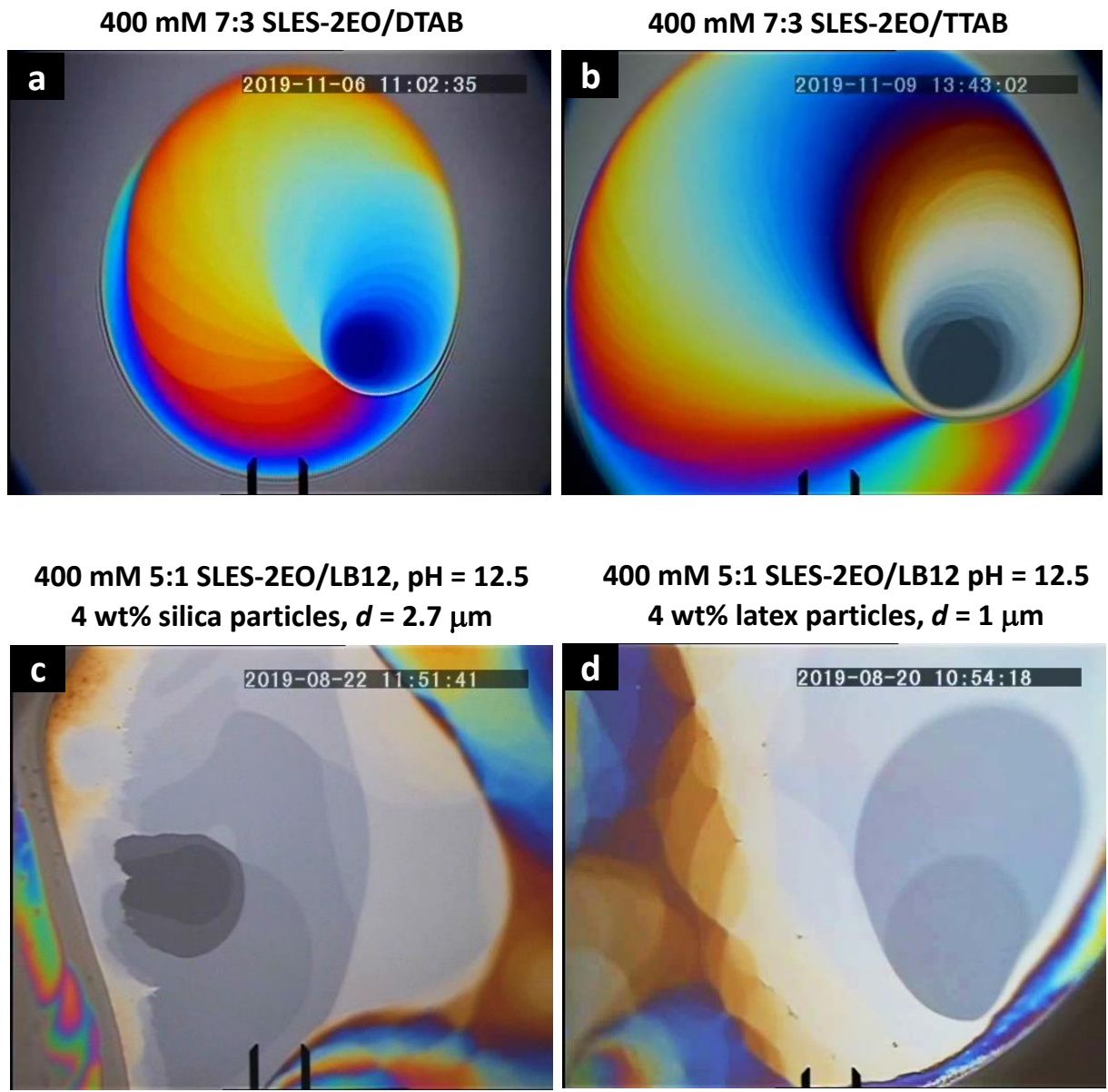

**Fig. S9. Observations of vortex with <u>other</u> systems:**
(a,b) In free liquid films from *turbid* solutions of other catanionic surfactant mixtures:
(a) 400 mM 7:3 SLES-2EO/DTAB (DTAB = dodecyltrimethylammonium bromide;
(b) 400 mM 7:3 SLES-2EO/TTAB (TTAB = tetradecyltrimethylammonium bromide. Stepwise profile is observed up to the crimson zone, which corresponds to film thickness ≈370 nm; see Table S1.
(c,d) In free liquid films from *clear* micellar solutions of 400 mM 5:1 SLES-2EO/LB12 at pH = 12.5 with added colloidal particles:
(c) 4 wt% silica particles of mean diameter $d = 2.7$ μm;
(d) 4 wt% latex particles of mean diameter $d = 1$ μm.

**Movie S1 (separate file)**

The movie shows the vortex appearance and evolution in an evaporating foam film formed in a SE cell from a solution of 400 mM 3:2 SLES-1EO / LB12 at pH = 6 – observation in reflected white light; the reference distance is 50 µm. First, the formation of thick colored film due to the penetration of liquid through the film periphery is observed. Next, the thick film of uneven thickness (and color) gradually occupies the whole film area and shows a fine structure of parallel stripes (stepwise thickness profile). Furthermore, one observes the stage of perfect vortex with a spot of thinner film in the center and stripes of different color (and thickness) around it. With time, the area of the central spot shrinks; its silvery white color indicates film thickness of ca. 120 nm. Furthermore, the thickness of the film increases – the yellow color of the central spot shows local film thickness of ca. 320 nm. Next, the central spot acquires orange color corresponding to thickness of ca. 350 nm. In this way, the thickening of the central (the thinnest) part of the film leads to self-protection of the film against breakage.

**Movie S2 (separate file)**

The movie shows the process of vortex disappearance after closing the experimental SE cell. The film is formed from a solution of 400 mM 3:2 SLES-1EO + LB12 with added 100 mM NaCl at pH = 6; see also Fig. S3. Upon gradual saturation of the water vapors in the closed cell and ceasing the evaporation from the film, the vortex disappears in the opposite way of its appearance. The process ends with the formation of a plane parallel film, which was initially observed before the opening of the experimental cell for evaporation. Reference distance = 50 µm

**Movie S3 (separate file)**

The movie shows the vortex appearance and evolution in an evaporating foam film formed in a SE cell from a solution of 400 mM 7:3 SLES-2EO / DTAB at the solution's natural pH – observation in reflected white light; the reference distance is 50 µm. The initial uniform film of magenta color has thickness of ca. 190 nm. Then, the experimental cell is opened to allow water evaporation and the film evolution is similar to that shown in Movie S1, with the main difference that the vortex develops with a thicker central spot, as indicated by the interference colors.